# Hard physics in PHENIX

**Dmitri Peressounko**
*RRC "Kurchatov Institute"*
*Kurchatov sq. 1, Moscow, 123182, Russia*
*E-mail:* `peressou@rcf.rhic.bnl.gov`

**for the PHENIX collaboration**[*]

We review recent results on hard observables in p+p, d+A and A+A collisions obtained by the PHENIX experiment. Emphasis is put on those measurements that provide insight into the properties of hot QCD media expected to be created in nucleus-nucleus collisions at RHIC energies. Direct photon spectra, jet properties and heavy quarks production measured in p+p and d+Au collisions are compared to the same observables extracted in heavy ion collisions to find modifications due to the presence of hot QCD matter.

---

[*] For the full list of PHENIX authors and acknowledgements, see [9].

One of the most important discoveries made since the start of the Relativistic Heavy Ion Collider was observation by four RHIC experiments of the suppression of the yield of high $P_t$ particles in Au+Au collisions comparing to binary scaled p+p collisions at the same energy and absence of such suppression in d+Au collisions [1-4]. These experiments demonstrated that a dense hot matter is indeed created in Au+Au collisions and it strongly influences partons, traversing it. Later a lot of efforts were devoted to detailed study of the properties of this matter. Below we outline some of the recent PHENIX results in this field.

It is straightforward to start analysis of the jet modification in Au+Au collisions looking at the angular distribution of particles in jet. Although a full reconstruction of a jet is not possible in heavy ion collisions due to hadron background in the underlying event, we studied jet shape modification by constructing angular correlation between a hard "trigger" hadron and a soft "associated" hadrons. As far as hadrons in the underlying event also carry angular correlations due to an elliptic flow, the resulting correlation will contain contributions from the jet and from the flow. We have separately measured strength of the elliptic flow and subtracted its contribution. The result of this procedure is presented in the Fig.1. For the most peripheral bin (plot f) there are clear near and away side peaks, corresponding to particles from same jet and opposite jet. However, for the 60% and more central collisions the away peak becomes wider and its shape changes so that in the most central collisions, we find "volcano-like" shape of the away peak. The reason of such a modification of the away jet shape in AA collisions could be such exiting effects as Mach cone [5], Cerenkov-like radiation [6] or something less exotic.

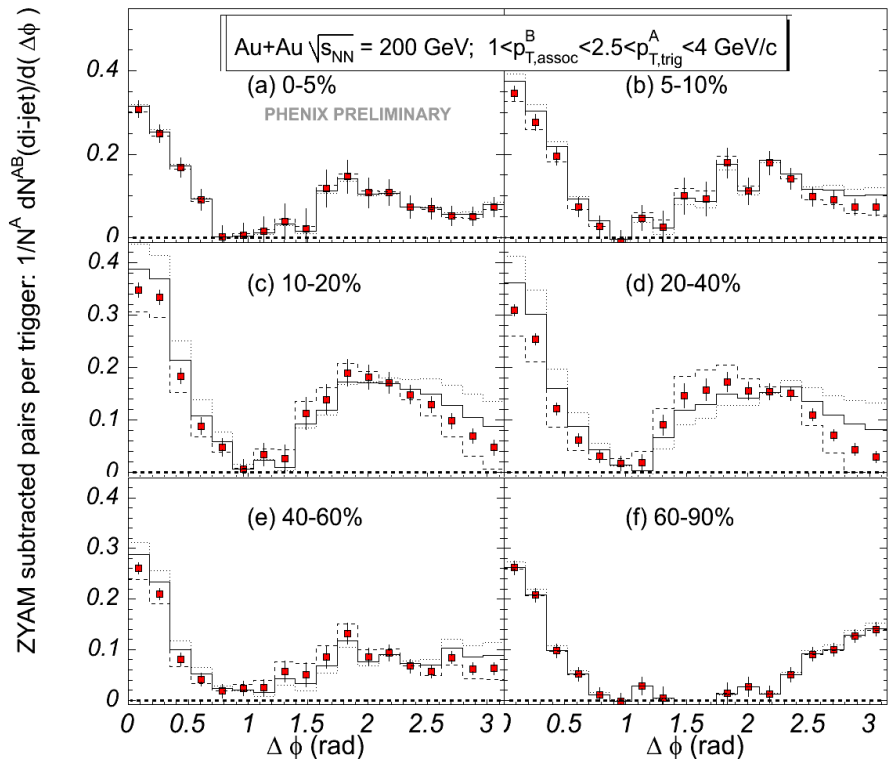

Figure 1. Number of associated particles per trigger as a function of azimuthal angle, measured in Au+Au collisions at $\sqrt{s_{NN}}$=200 GeV at different centralities.

To make any *quantitative* conclusion concerning modification of the jet properties, one first should have full control over the initial state of the collision. One of the best ways to test initial state in AA collisions is the measurement of the direct photon yield. Having large free path length, direct photons escape from the hot matter without rescattering and provide clear information about the initial state. Direct photons, emitted in the collisions of quarks of the colliding nuclei dominate in high $P_t$ region, while photon emission from the later stages, such as thermal emission of the hot matter, appears at intermediate and soft $P_t$. To provide a baseline for the direct photon emission in Au+Au collisions, we measured direct photon yield in p+p and d+Au collisions at $\sqrt{s_{NN}}$=200 GeV and compared it to the NLO pQCD predictions [7], scaled with number of binary collisions (Fig.2, left plot). Three curves correspond to three choices of renormalization and factorization scales. Comparing our data to theory, we find good agreement both in p+p and d+Au reactions, in contrast to lower $\sqrt{s}$ energies, where one has to introduce some "$K_T$ broadening" to reproduce the data. Such an agreement means that modifications of structure functions in nuclei, such as shadowing and Cronin effect, if present at this $x_B$ range, are rather modest in these collisions and do not exceed 10-20%. Finally, we measured direct photon yield in Au+Au collisions for several centrality classes [8]. For all centralities data exhibit scaling with number of binary collisions and agree with the pQCD calculations. So, we can conclude that there is no considerable modification of the *initial* state in Au+Au collisions. The statistical and systematic errors at small transverse momenta ($P_t$<4-5 GeV) are still too large to make any definite conclusion, but we do find some enhancement with respect to pQCD line, consistent with thermal photon emission.

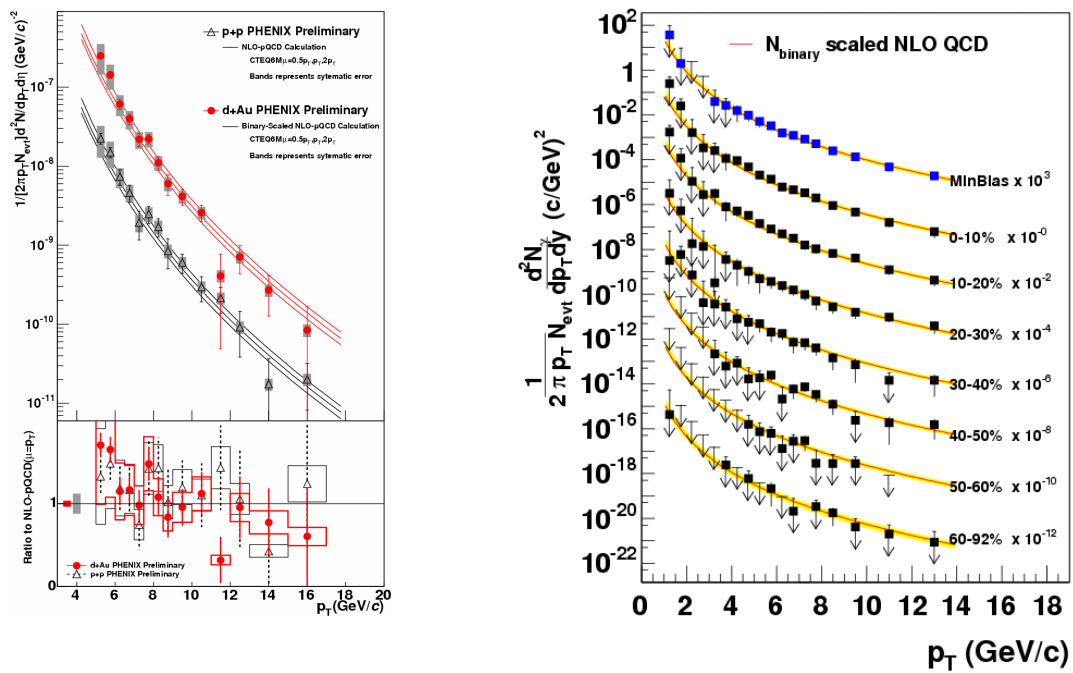

Figure 2. Direct photon yield in p+p, d+Au (left plot) and in Au+Au (right plot) collisions at $\sqrt{s_{NN}}$=200 GeV, compared to NLO pQCD predictions.

Another complimentary way to study properties of the hot matter, created in the nucleus-nucleus collisions is production of heavy quarkonia. Since the energy loss by heavy quark, traversing QGP is under much better control than in the case of light quarks, this analysis provides important complimentary information. PHENIX studied open charm production using its semileptonic decays. In particular, we subtract from inclusive single electron spectrum contributions from neutral pions Dalitz decays, photon conversion, light mesons decays, direct radiation and Kaons weak decays. The rest - "single non-photonic electrons" - come from charm and bottom decays. We find that in p+p collisions yield is slightly harder than PYTHIA predicts. Measurements made in d+Au collisions demonstrate good agreement with p+p data, scaled with number of binary collisions (see Fig.3, left plot), suggesting absence of the considerable modifications in gluon distributions in nuclei. We measured non-photonic single electron spectra in Au+Au collisions in several centrality classes and compared them to binary scaled spectrum in p+p collisions. We found reasonable agreement in the soft region, while at high $P_t>2.5-5$ GeV there is some suppression (Fig.3, right plot). Quantitative analysis shows that this suppression is consistent with neutral pion suppression and thus the hot matter is so dense that even heavy quarks are stopped.

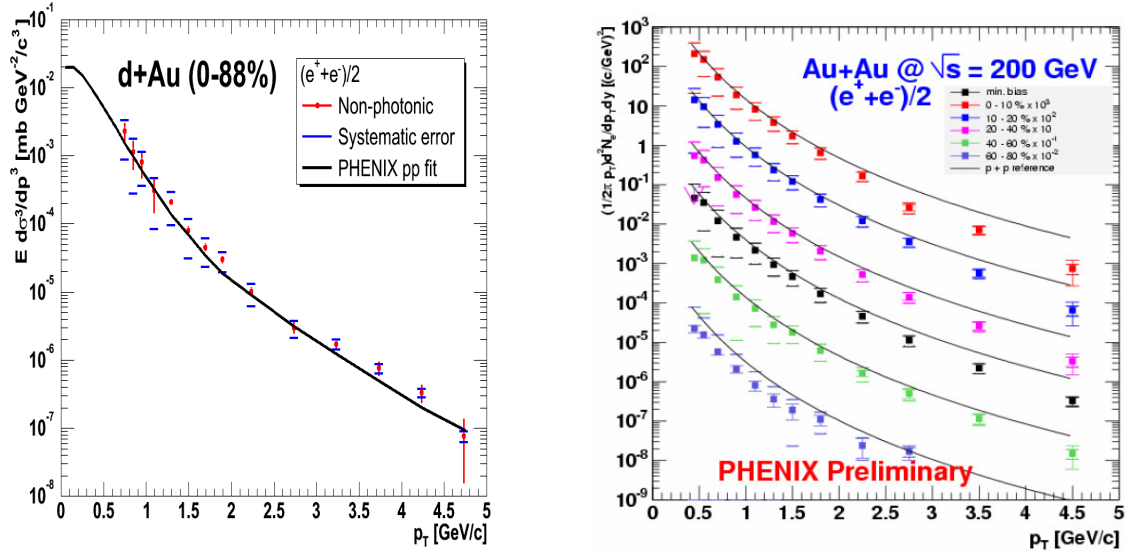

Figure 3. Single non-photonic electron yield in d+Au and Au+Au collisions, compared to fit to yield in pp.

To conclude, PHENIX experiment explored properties of the hot matter, created in the heavy ion collision, using very different observables: we tested initial state of the collision with direct photons, measured modification of the jet shape in AA collisions with hadrons and explored energy loss of heavy quarks using lepton production. All this, being small part of all PHENIX results [9], provides us with new exciting information about properties of hot matter.

**References**

[1] S. S. Adler *et al.* (PHENIX Collaboration), Phys.Rev.Lett.91, (2003) 072301, ibid 072303.
[2] B. B. Back *et al.* (PHOBOS Collaboration), Phys.Rev.Lett.91, (2003) 072302.
[3] J. Adams *et al.* (STAR Collaboration), Phys.Rev.Lett.91, (2003) 072304.
[4] I. Arsene *et al.* (BRAHMS Collaboration), ), Phys.Rev.Lett.91, (2003) 072305.
[5] J. Casalderrey-Solana et al., hep-ph/0411315.
[6] A. Majumder and X.-N. Wang, nucl-th/0507062.
[7] L.E.Gordon and W.Vogelsang, Phys.Rev.D 48 (1993)3136; Phys.Rev. D 50 (1994)1901.
[8] S.S.Adler et al, (PHENIX collaboration), Phys. Rev. Lett. 94, 232301 (2005).
[9] K. Adcox, et al. (PHENIX collaboration), Nucl. Phys. A757 (2005) 184.